\documentclass{PoS}

\title{In memory of Mikhail Igorevich Polikarpov}

\ShortTitle{In memory of M.I. Polikarpov }

\author{\speaker{V.~G.~Bornyakov}\\
        Institute for High Energy Physics, 124281 Protvino, Russia\\
        Institute of Theoretical and Experimental Physics, 117259 Moscow, Russia\\
        E-mail: \email{vitaly.bornyakov@ihep.ru}}

 \author{P.~V.~Buividovich$^*$\\
        Institute for Theoretical Physics, Regensburg University, Universitatsstrasse 31, D-93053
Regensburg, Germany \\
        E-mail: \email{pavel.buividovich@physik.uni-regensburg.de}}

\author{M.~A.~Zubkov\\
        Institute of Theoretical and Experimental Physics, 117259 Moscow, Russia\\
        The University of Western Ontario, Department of Applied Mathematics,
 1151 Richmond St. N., London (ON), Canada N6A 5B7\\
        E-mail: \email{zubkov@itep.ru}}

\abstract{\vspace*{-0.5cm}
This obituary is devoted to M.~I.~Polikarpov (28.12.1952 - 18.07.2013), our teacher, our colleague, our friend. We recollect some facts of his biography and stages of his scientific career, and make a brief review of some of his most known scientific works. We conclude with some messages of condolence which were received from his colleagues and friends after his sudden death.\\
\begin{center}
\vspace*{-0.5cm}
\hspace{2cm}
\includegraphics[angle=0,scale=0.30]{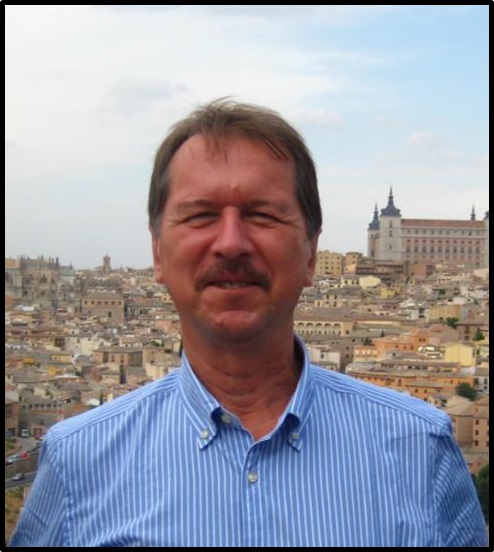}
\vspace*{-5.3cm}
\end{center}
}

\FullConference{31st International Symposium on Lattice Field Theory LATTICE 2013\\
                 July 29 $-$ August 3, 2013\\
                 Mainz, Germany}

\begin{document}

\section*{Facts of biography}

 Mikhail Igorevich Polikarpov, Misha for his friends and Western colleagues, was born on Dec. 28, 1951. Everybody who knew Misha could freely come to visit him on his birthday. In early years of his childhood Misha spent a few years in France together with his parents. Back in Moscow he attended special school with an extensive study of French.
He went to study at the Department of General and Applied Physics of the Moscow Institute for Physics and Technology (MIPT) in 1969. Misha joined group number 27 with specialization in elementary particle physics. The competition was unthinkable - more than 100 applicants per one place in the group. His university mates called him Michelle because of his excellent French.  Already at the second year the students at MIPT attended lectures at ITEP. Prof. Yuri Simonov became Misha's supervisor for both diploma thesis and PhD thesis.

\noindent
``{\it From the very beginning Misha was looking for new, original ways to solve the problems which I suggested him.    One example was use of Green function Monte Carlo method. He moved to lattice gauge theory studies soon after getting PhD, that was again his own choice. I believe he made a good service for the Russian science, without him the development of the important  field of lattice gauge theory would be delayed in Russia by many years.}'' \hfill                   {\bf  Yu.~Simonov    }

 Misha was working at ITEP since 1975, for almost 40 years. He started research in the field of lattice gauge theory in 1980 together with Yuri Makeenko and late Sasha Veselov. Lattice ITEP group was formally created in 2001 with the name ``Laboratory 191 on Lattice gauge theories''.

 The list of Misha's collaborators is very impressive. HepNames gives more than 90 names, among them more than 30 foreign colleagues. His most frequent collaborator is one of his best students - Maxim Chernodub, now at CNRS in Tour, France. They wrote together 83 papers. Misha had a special talent to attract young talented students. His mission was to teach them, to help them to become excellent scientists.

 Misha loved his family very much. He left two sons. Sergey (20) is a 5th year student at MIPT and a member of LHCb collaboration.  Andrey (18) is a 4th year student at the department of Computational Mathematics and Cybernetics of the Lomonosov Moscow State University.

\section*{Scientific career}

The first published paper \cite{Badalian:1975pa} by Mikhail~Polikarpov is dated 1975. It was devoted to the general properties of amplitudes in hadronic physics. Within the period between 1975 and 1982 more than 20 papers were published on the similar subjects in collaboration with Yu.~A.~Simonov and A.~M.~Badalyan. The review of the results obtained during this period appeared in Physics Reports \cite{rept1} in 1982.

 Mikhail started to study lattice gauge theories in collaboration with Yu.~M.~Makeenko and A.~I.~Veselov \cite{Makeenko:1981bb, Polikarpov:1987yr}. In particular, in \cite{Polikarpov:1987yr} the results of the first numerical investigation of instantons in gluodynamics were presented. The properties of gauge field configurations obtained by cooling the quantum fluctuations of the gauge fields on the lattice were investigated. It was found that the cooling process can be subdivided into three distinct stages: the emergence of semi-classical instanton-anti-instanton vacuum, annihilation of instantons and anti-instantons, and, finally, relaxation to stable classical solutions. The distributions of instanton radii and inter-instanton distances were obtained. It was demonstrated that the contribution of the multi-instanton configurations to the value of the string tension is about $5\%$.

 Remarkably, the analytical multi-instanton solution was also obtained by M.~I.~Polikarpov for the first time. This result remained unpublished and we became aware of it only after his death. A.~A.~Belavin (who was one of the authors of the first paper on instantons \cite{Belavin:1975fg}) told us this story. The authors of \cite{multiinstantons1} derived the equations that define the multi-instanton solutions in the Yang-Mills theory. M.~I.~Polikarpov was the first who obtained the analytical solutions of these equations. This was mentioned in \cite{multiinstantons1}, where these solutions were presented. It is worth mentioning, that independently and simultaneously the same problem was solved by E.~Witten \cite{multiinstantons2}.

 Prof.~M. Polikarpov started to create his own scientific group in 1990. The first paper published by this group \cite{Ivanenko:1990xu} was written in collaboration with his graduate students T.~L.~Ivanenko and A.~V.~Pochinsky. In this paper it was numerically demonstrated in $SU(2)$ lattice gauge theory in maximal Abelian gauge that the density of Abelian monopoles which occupy several lattice cells (extended monopoles) is strongly correlated with the string tension. This result indicates that the string between quark and antiquark is the dual analog of the Abrikosov vortex in superconductors. In this study for the first time the Hausdorf dimension of the monopole current line was used as an order parameter.

 After M.~I.~Polikarpov received his Habilitation degree in 1991, one of us (M.~A.~Z.) became his graduate student. In 1993 the paper \cite{Polikarpov:1993cc} was written, in which the partition function of the 4D lattice Abelian Higgs theory is represented as the sum over world sheets of Nielsen-Olesen strings. The creation and annihilation operators of these strings were constructed. The topological long-range interaction of the strings and charged particles was shown to exist; it is proportional to the linking number of the string world sheet and particle world trajectory.

 In the later years Emil Akhmedov, Maxim Chernodub, and Fedor Gubarev joined the group of Prof.~Polikarpov. As well as M.~A.~Z. they first became graduate students, and then PhD students of M.~I.~Polikarpov. Some of the most known works done in collaboration with Fedor and Maxim are \cite{Chernodub:1996ps,Chernodub:1997dr}, where the authors have performed numerical calculation of the probability distribution of the value of the monopole creation operator in $SU(2)$ lattice gauge theory. It was found that in the low-temperature confining phase the maximum of this distribution is shifted from zero, which means that the effective constraint potential is of the Higgs type. Above the phase transition the minimum of the potential (the maximum of the monopole field distribution) is at the zero value of the monopole field. This fact confirms the existence of the Abelian monopole condensate in the confinement phase of lattice gluodynamics, and agrees with the dual superconductor model of the confining vacuum.

 Emil Akhmedov became an independent researcher almost immediately after getting his PhD. Nevertheless, his own line of research began from the work \cite{Akhmedov:1995mw} written under the supervision of M.~I.~Polikarpov. In this paper the continuum theory of quantum Abrikosov-Nielsen-Olesen strings was constructed starting from the Abelian Higgs field theory. It was shown that in four space-time dimensions in the limit of infinitely thin strings, the conformal anomaly is absent, and the quantum theory of such strings exists.

 The new impulse was given to the ITEP lattice group headed by Misha Polikarpov when it was joined by the world famous physicist V.~I.~Zakharov. His experience and scientific intuition became the important ingredient of the further research work carried out by the group. With his participation, for example, the paper on anatomy of Abelian monopoles appeared \cite{Bornyakov:2001hs}. In this paper the Abelian and non-Abelian action densities near the monopole in the maximal Abelian gauge of SU(2) lattice gauge theory were studied. It was found that the non-Abelian action density near the  monopoles belonging to the percolating cluster decreases when the monopole center is approached. The estimate of the monopole radius was given ($\approx 0.04$ fm). Approximately at that time one of us (V.~G.~B.) joined ITEP lattice group as well.

 The next generation of the members of the ITEP lattice group are A.~V.~Kovalenko, S.~N.~Syritsyn, S.~M.~Morozov and P.~Yu.~Boyko, who were the graduate students at ITEP. Later S.~N.~Syritsyn moved to US, while A.~V.~Kovalenko, S.~M.~Morozov and P.~Yu.~Boyko became PhD students and then researchers at ITEP. Around that time also V.~A.~Belavin became the PhD student of M.~I.~Polikarpov. After getting his PhD degree he moved to the investigation of conformal field theories. In the series of papers written by M.~I.~Polikarpov in collaboration with V.~G.~Bornyakov,M.~N.~Chernodub, F.~V.~Gubarev, V.~I.~Zakharov, A.~V.~Kovalenko, S.~N.~Syritsyn, S.~M.~Morozov and P.~Yu.~Boyko the lattice studies of the structure of QCD vacuum were continued. Some of the well-known publications of this period are \cite{Gubarev:2002ek,Bornyakov:2003ee}. In \cite{Gubarev:2002ek} measurements of the action associated with center vortices in $SU(2)$ pure lattice gauge theory were performed. In lattice units the excess of the action on the plaquettes belonging to the vortex turned out to be approximately constant and independent of the lattice spacing $a$. Therefore, the action of the center vortex is of order $A/a^2$, where $A$ is its area. Since the area $A$ is known to scale in the physical units, the measurements imply that the suppression due to the surface action is balanced, or fine tuned to the entropy factor which is to be an exponential of $A/a^2$.

 In \cite{Bornyakov:2003ee} propagators of the diagonal and the off-diagonal gluons were studied numerically in the maximal Abelian gauge of $SU(2)$ lattice gauge theory. It was found that in the infrared region the propagator of the diagonal gluon was strongly enhanced in comparison with the off-diagonal one. The enhancement factor is about $50$ at the smallest momentum $325$ MeV.

 Around 2007 one of us (P.~V.~B.) and Elena Luschevskaya joined the ITEP lattice group. Both were the PhD students of M.~I.~Polikarpov, and then researchers at ITEP. At this time the research direction of the group shifted to the lattice studies of the effect of strong magnetic fields on QCD vacuum. In particular, in \cite{Buividovich:2009wi} the microscopic picture of the Chiral Magnetic Effect was studied on the lattice. There is a recent evidence that this effect is observed by the STAR Collaboration in heavy ion collisions at RHIC. In \cite{Buividovich:2009wi} it was found that indeed magnetic field strongly enhances the fluctuations of electric current in the direction of the magnetic field. In \cite{Buividovich:2010tn} this enhancement was related to the emergence of electric conductivity in the direction of magnetic field. It was predicted that this phenomenon should manifest itself in specific anisotropy of the distribution of soft leptons produced in heavy-ion collisions.

The last scientific topic investigated by Misha Polikarpov was the physics of graphene. In \cite{Ulybyshev:2013swa} a first-principle numerical study of spontaneous breaking of chiral (sublattice) symmetry in suspended monolayer graphene due to electrostatic interaction was performed. For the first time the screening of Coulomb potential by electrons on $\sigma$-orbitals was taken into account. In contrast to the results of previous numerical simulations with unscreened potential, it was found that suspended graphene is in the conducting phase with unbroken chiral symmetry. This finding is in agreement with the most recent experimental results obtained by the group of K.~Novoselov and A.~Geim.

M.~I.~Polikarpov collaborated fruitfully with the foreign colleagues from many countries. His long-term collaborators were T.~Suzuki, M.~M\"uller-Preussker, G.~Schierholz, M.~Ilgenfritz, G.~Greensite, B.~Bakker.

\vspace{.5cm}
 To conclude, let us simply list the former and current members of the ITEP lattice group, which was created by Misha Polikarpov and which is still the only lattice group in Russia:
Emil Akhmedov,
Vladimir Belavin,
Vitaly Bornyakov,
Pavel Boyko,
Victor Braguta,
Pavel Buividovich,
Maxim Chernodub,
Fedor Gubarev,
Taras Ivanenko,
Oleg Kochetkov,
Anton Kononenko,
Andrey Kotov,
Alexey Kovalenko,
Olga Larina,
Elena Luschevskaya,
Yurij Makeenko,
Boris Martemyanov,
Valentin Mitrjushkin,
Alexander Molochkov,
Sergey Morozov,
Oleg Pavlovsky,
Andrey Pochinsky.
Mikhail Prokudin,
Roman Rogalev,
Vladimir Shevchenko,
Sergey Syritsyn,
Maxim Ulybyshev,
Semen Valgushev,
Alexander Veselov,
Valentin Zakharov,
Mikhail Zubkov.

 The people listed above are the former students of Misha Polikarpov, his colleagues and friends. Not all members of the lattice ITEP group continue now to work in science. But, no doubts, all of them experienced the unique atmosphere of the scientific group created with love and passion by Misha Polikarpov and all of them will keep him in their hearts and memories forever.

\section*{Messages of condolence from colleagues}

\noindent
``{\it For us, his students, PhD students and PostDocs, Misha Polikarpov was like a scientific father.
And much more than just that, his care about us was very personal and very kind.
He has deeply influenced many of us. And it is very hard to believe that we will not see
Misha anymore ... at least in this world.}''
\hfill                   {\bf Maxim Chernodub}

\vspace{0.2cm}
\noindent
``{\it Professional activity of Mikhail Polikarpov is well known to the participants of the Conference.
What was, probably, clearer seen from close distance is that this activity was based on his
moral principles which determined his role as a scientist, teacher, citizen, believer.
This explains, in particular, his devotion to going to ever changing frontiers in science, taking care
of everyday-life problems of his group, staying in Russia through all these years, and so on.}''
\hfill                                                                 {\bf Valentin   Zarharov}

\vspace{0.2cm}
\noindent
``{\it I knew Mikhail since many years.  He always impressed me by his optimism and
engagement to pick up new interesting problems to be solved with lattice methods: e.g.
studying the Yang Mills vacuum state with the overlap operator, discussing the entanglement
entropy, computing properties of gluon-quark matter in an external magnetic field or in the
last period investigating graphene with different lattice geometries. But most important
 I consider his role to create and maintain a Russian center of lattice field theory and his promotion
of many talented young people under the difficult Russian conditions.}''
\hfill    {\bf Michael Mueller-Preussker}

\vspace{0.2cm}
\noindent
``{\it Misha was really a wonderful and attractive person with a gentle mind. He was literally one
of great leaders in the world lattice community. His works on topological properties of non-perturbative
QCD are world famous and I am very much grateful to him for having able to share a small part
of them as an ITEP-Kanazawa collaboration. I believe our long-term collaboration was very fruitful
and successful mainly due to Misha$'$s warm and active endeavor. I and many young colleagues of
mine are very much obliged to him. In addition to his own scientific works, it is really great that Misha
has developed a big nice active lattice group in Russia, raising many brilliant young scientists who
are now world-wildly working very actively.}''
\hfill                                                                {\bf Suzuki Tsuneo}

\vspace{0.2cm}
\noindent
``{\it I knew Misha for more than 20 years. He was an amazing person.
He founded a highly respected lattice group in Russia, all by himself, carrying the ITEP spirit.
His students were world class, some of them made it to the US and elsewhere. And last not least,
he converted a barn into a respectable office building on the ITEP site, equipped with computers and printers.}'' \\
\hspace*{12.1cm}                                              {\bf Gerrit Schierholz}

\vspace{0.2cm}
\noindent
 ``{\it Some years ago people were hotly debating the nature of confining configurations in QCD.
Misha Polikarpov and I were often on opposite sides of that discussion.
You can learn a lot from debating with a worthy and honorable opponent, and that's the kind
of opponent that Misha was. He wasn't interested in scoring points in some scientific contest;
his interest was in getting to the bottom of things.  Later on Misha and I became collaborators
on a number of scientific projects, we also worked together to organize the confinement sessions
at the Confinement meetings, and I had a chance to observe Misha's role as a mentor to the
younger scientists at ITEP. It was a privilege to work with him.  Misha's premature death
came as shock to me, and to many of us.  He will be sorely missed in our community.}''
\hfill                                                             {\bf  Jeff Greensite}

\vspace{0.2cm}
\noindent
 ``{\it Misha Polikarpov created and nurtured the Russian Lattice group.
    He was always open-minded about new ideas, curious and ready to investigate.
He kept us young, and we will miss him.}'' \\
\hspace*{11.6cm} {\bf Philippe de Forcrand   }

\vspace{0.0cm}
\noindent
``{\it I always admired Misha both as scientist and man, looked forward to meeting and
discussing with him. I will miss him a lot, and can imagine how much he will be missed
by all his colleagues, collaborators and students.}''
\hfill {\bf  Stefan Olejnik}

\vspace{0.2cm}
\noindent
``{\it Needless to say that this is very sad news indeed. I still cannot
believe it. I have no idea how lab 191 will continue without such an
organizational talent.}'' \hfill {\bf Gunnar Bali }

\vspace{0.2cm}
\noindent
 ``{\it The news came like a great blow to me. Only a few weeks ago I was
dining and talking a lot with Misha in Benasque. I had the highest consideration
for him as a scientist and a human being. I feel that I have lost a friend. My last preprint
is dedicated to his memory.}''\hfill {\bf Tony Gonzalez-Arroyo}

\vspace{0.2cm}
\noindent
``{\it  This is indeed very shocking and tragic news. The sudden death of Mikhail marks not
only the loss of a friend but also that of a great scientist. Mikhail was truly instrumental
in setting up the lattice group at ITEP, which has produced many outstanding and talented physicists.
 He will be sorely missed.}''\hfill {\bf Hartmut Wittig}

\vspace{0.2cm}
\noindent
``{\it It is hard to express the astonishment and the sadness for such news.
This is a major loss for the ITEP group, but it is also a major loss for the whole lattice community.}'' \hfill         {\bf  Massimo D'Elia}

\vspace{0.2cm}
\noindent
``{\it Misha was one of my longest time friends from Russia, and one of my hosts
when Alice and I first visited Moscow and ITEP. Misha I know was a believer, even
during the early times  when it was difficult. In Spanish they have an expression,
when you go away on a trip, "Vaya con dios" which simply means "Go with God". Perhaps he will. I hope so.}''
\hfill                                {\bf Larry McLerran}

\vspace{0.2cm}
\noindent
``{\it Dear friends of the ITEP group, let me join you in remembering a friend and an outstanding colleague.
 May be the best way to honour him is to continue what he had started.}''\\
\hspace*{11.6cm} {\bf Adriano Di Giacomo}

\vspace{0.2cm}
\noindent
 ``{\it The tragic news. A huge loss for the ITEP for science. A huge loss for Russian participation in the FAIR project.}''
\hfill
                                                                                       {\bf Boris Sharkov}
\vspace{0.2cm}
\noindent
 ``{\it Shocked by the sudden death of Mikhail I. Polikarpov, outstanding  scientist
and remarkable person.}'' \hspace{12.5cm}{\bf Lev Okun}

\vspace{0.2cm}
 \noindent
 ``{\it I heard his brilliant report a few days ago at a conference of the Institute of Euler
in St. Petersburg and I can not believe that I will not meet Misha to discuss the current research issues.
He will always be remembered as an outstanding scientist and a wonderful person.}''\hfill                                                                                        {\bf Lev Lipatov}

\end{document}